\documentclass[a4paper,superscriptaddress,aps,prl,twocolumn,floatfix,citeautoscript]{revtex4-1}
\usepackage{amsfonts,amssymb,amsmath,latexsym}
\usepackage{xcolor}
\usepackage{setspace}
\usepackage{graphicx}
\usepackage{multirow}
\usepackage{tabularx}
\usepackage{float}
\usepackage{bigstrut}
\usepackage{caption}
\usepackage{subcaption}
\usepackage{booktabs}
 \usepackage[margin=0.80in]{geometry}
\usepackage[normalem]{ulem}

\newcommand{\lcpq}{Laboratoire de Chimie et Physique Quantiques, CNRS, Universit\'e de Toulouse, France}
\newcommand{\etsf}{European Theoretical Spectroscopy Facility (ETSF)}
\newcommand{\unibo}{Universit\`a di Bologna, Bologna, Italy}
\newcommand{\unife}{Dipartimento di Scienze Chimiche, Farmaceutiche ed Agrarie, Universit\`a di Ferrara, Ferrara, Italy}
\newcommand{\uniga}{Islamic University of Gaza, P.O. Box 108, Gaza, Palestine}

\begin{document}
\title{ 
A unique one-body position operator for periodic systems
}

\author{Stefano Evangelisti}
\email{stefano.evangelisti@irsamc.ups-tlse.fr}
\affiliation{\lcpq}
\author{Faten Abu-Shoga}
\affiliation{\uniga}
\author{Celestino Angeli}
\affiliation{\unife}
\author{Gian Luigi Bendazzoli}
\affiliation{\unibo}
\author{J.~Arjan Berger}
\email{arjan.berger@irsamc.ups-tlse.fr}
\affiliation{\lcpq}
\affiliation{\etsf}


\date{\today}

\begin{abstract}
In this work we prove that the one-body position operator for periodic systems that we have recently proposed [Phys. Rev. B 99, 205144] is unique modulo a phase factor and an additive constant.
The proof uses several general physical constraints that a periodic one-body position operator should satisfy.
We show that these constraints are sufficient to uniquely define a position operator that is compatible with periodic boundary conditions.
\end{abstract}

\maketitle

The position of a particle is a fundamental concept in physics.
In classical mechanics the position is a vector whose components are the Cartesian coordinates of the particle,
while in quantum mechanics the position is an operator whose action is the multiplication by the coordinates of the particle.
However, these definitions implicitly assume that we work within open boundary conditions (OBC).
Instead, many systems, in particular crystalline solids, are more efficiently described within periodic boundary conditions (PBC).
In fact, the use of PBC is a crucial tool for the treatment of crystals and periodic systems in general.
Unfortunately, the above definitions are not compatible with PBC since they do not yield a unique definition of the position.

Some solutions have been proposed in the literature.
Zak proposed to define the position operator in terms of a sine function~\cite{Zak_2000}.
However, this proposition yields a distance function that depends on the choice of the origin of the coordinate system, and this is unphysical.
It has been shown that discontinuous position functions, such as the sawtooth function, are also problematic.
For example, they cannot be used to extract the electronic response of periodic chains to a uniform external field~\cite{kirtman_2009}.
Finally, Resta proposed an expression for the expectation value of the total position operator 
that is compatible with PBC~\cite{Resta_1998}.
Besides the fact that it is not a definition of the position operator itself, the expression involves an explicit many-body operator, and therefore, becomes unpractical in studies of systems with many electrons.

Recently, we have proposed a periodic position $q_L$ that is compatible with PBC~\cite{Valenca_2019}.
In one dimension it is given by
\begin{equation}
q_L(x) = \frac{L}{2\pi i} \left[\exp\left(\frac{2\pi i}{L}x\right) -1\right]
\label{Eqn:qlx}
\end{equation}
where $L$ is the period.
We then demonstrated that $q_L(x)$ as defined above complies with several important constraints that should be satisfied by a periodic position.
We have successfully applied this position and the corresponding distance function in the calculation of Madelung constants in ionic crystals~\cite{Tavernier_2020,Tavernier_2021}, in the calculation of ground state energies of Wigner crystals~\cite{Alves_2021} and Wigner molecules~\cite{Escobar_2019,Escobar_2021}, as well as the localization tensors and polarizabilities of periodic chains~\cite{Valenca_2019,Angeli_2021}. Recently it has also been used in the calculation of the localization tensor of one-dimensional mosaic lattices~\cite{Gong_2021}.

In this work we will make an important further step and prove that, given the known constraints for the position operator within PBC, Eq.~\eqref{Eqn:qlx} is the only definition that satisfies these constraints. 
In other words, we proof the uniqueness of the periodic position in Eq.~(\ref{Eqn:qlx}).
We thus solve a longstanding problem of defining a position operator that is consistent with PBC.
For the sake of clarity we will focus here on the one-dimensional (1D) case. 
The generalization to many-dimensional systems is straightforward, provided the different dimensions are along mutually orthogonal axes.

Our strategy for the definition of a classical periodic position, and then to move to a quantum (possibly many-body) context, 
mimics the strategy in the case or ordinary OBC.
It is based on the following steps:
\begin{enumerate}
    \item 
    Define a classical periodic position suitable for the treatment of periodic systems.
    \item
    Define a quantum position operator as in the non-periodic case, \emph{i.e.}, a multiplicative operator, whose action is the multiplication by the classical periodic position.
    \item
    In the case of a many-body system, the total many-body operator is obtained by adding the individual one-body periodic positions of all the particles of the system.
\end{enumerate}
%

In the 1D case a periodic system is isomorphic to a circle.
Let us consider therefore the periodic interval $[0,L]$, where $L$ is the length of the period, and let $x$ be the coordinate of a point belonging to this interval, \emph{i.e,} $x \in [0,L]$.
Let us call the (possibly complex) function $q_L(x)$ the {\em periodic position} associated the coordinate $x$.
Because of periodicity $x$ is defined modulo the length $L$.
We impose the following four general and physically motivated conditions on the periodic position $q_L(x)$:
\begin{enumerate}
    \item 
    The periodic position $q_L(x)$ is invariant with respect to translations equal to the length $L$.
    In other words, $q_L(x)$ is a periodic function with period $L$, \emph{i.e.},
    \begin{equation}
        q_L(x+L)  =  q_L(x) \; \; \forall x  .
    \end{equation}
    \item
    The periodic position $q_L(x)$ is a \emph{simple} periodic function, \emph{i.e.}, 
    there is a one-to-one correspondence between $x$ and $q_L(x)$. 
    This condition guarantees that $L$ is the smallest period of $q_L(x)$, \emph{i.e.},
    \begin{equation}
        x\neq0 \; \; \; (\hspace{-4mm} \mod 2\pi) \; \; \Rightarrow \; \; q_L(x) \neq q_L(0)  .
    \end{equation}
    \item
    The distance between two arbitrary points $x$ and $x+d$ is the modulus of the difference between the corresponding periodic positions, $q_L(x)$ and $q_L(x+d)$, \emph{i.e.},
    \begin{equation}
        |q_L(x+d)-q_L(x)|  =  |q_L(d)-q_L(0)|  .
    \end{equation}
    This condition ensures that the distance between $q_L(x)$ and $q_L(x+d)$ is a function of $d$ alone, independent of $x$.
    %
    %
    \item
    In the limit of very large values of $L$, and for a {\em fixed} value of $d$, 
    we must obtain the OBC distance between the two points, \emph{i.e.},
    \begin{equation}
        \lim_{L \rightarrow \infty} \, |q_L(d)-q_L(0)|^2  =  d^2  .
        \label{Eqn:cond4}
    \end{equation}
    This implies that, in the limit of an infinite system, we must recover locally the non-periodic OBC distance.
\end{enumerate}
In the following we will show that the above four constraints on $q_L(x)$ are sufficient to uniquely define the periodic position given in Eq.~\eqref{Eqn:qlx}.

Condition 1 implies that we can express $q_L(x)$ as a Fourier series of period $L$ according to
%
\begin{equation}
    q_L(x)  =  \sum_{k=-\infty}^\infty a_k \, \exp \left(\frac{i 2\pi kx}{L}\right) \; ,
    \label{Eqn:Fourier}
\end{equation}
where $a_k$ are complex coefficients.
We note that the functions $\exp(i2\pi kx/L)$ for different values of the integer $k$ are orthogonal and, therefore, linearly independent, on the interval $[0,L]$. 
This fact will play a crucial role in the following.

In order to impose Condition 3 we first compute the square modulus of the difference $D_L(x,d) \, \equiv \, q_L(x+d) \, - \, q_L(x)$ using Eq.~\eqref{Eqn:Fourier}. We obtain
%
%
\begin{equation}
    |D_L(\bar{x},\bar{d})|^2 \!=\!\! \sum_{k,l=-\infty}^{\infty} \!\!\!\! a_k^* a_l 
    e^{i(l-k)\bar{x}}
    \left[e^{-i k\bar{d}}-1\right] \left[ e^{i l \bar{d}}-1\right],
    \label{eq:dist2}
\end{equation}
in which we defined the scaled quantities $\bar{x}=2\pi x/L$ and $\bar{d}=2\pi d/L$ and where $a_k^*$ is the complex conjugate of $a_k$.
If the square modulus of $D(\bar{x},d)$ is independent of $\bar{x}$, 
its derivative with respect to $\bar{x}$ must be identically zero.
Therefore, we can impose Condition 2 with the following relation,
\begin{align}
\notag
    \frac{\partial}{\partial \bar{x}} 
    |D(\bar{x},\bar{d})|^2 \! & =\!\! 
    \sum_{k,l=-\infty}^{\infty} \!\!\!\! i a_k^* a_l (l-k) e^{i(l-k) \bar{x}}
    \\ & \times
    \left[e^{-i k\bar{d}}-1\right] 
    \left[e^{i l\bar{d}}-1\right] = 0  .
    \label{eq:dist_der}
\end{align}
We now make two changes of variable, $r=l+k$ and $s=l-k$, 
which imply $l=(r+s)/2$ and $k=(r-s)/2$.
We note that $r$ and $s$ are either both even or both odd, \emph{i.e.}, $r\equiv s (\mathrm{mod}\,2)$.
The above equation can thus be rewritten as
\begin{align}
\notag
    & \sum_{s=-\infty}^{\infty}
    \sum_{\substack{r={-\infty} \\ r\equiv s (\mathrm{mod}\,2)}}^{\infty}
    i s a_{(r-s)/2}^* a_{(r+s)/2} e^{is \bar{x}} 
    \\ &\times 
    \left[e^{-i (r-s)\bar{d}/2}-1\right] \left[e^{i (r+s)\bar{d}/2}-1\right] = 0 .
    \label{eq:s_sr}
\end{align}
%
We now write the above equation according to
\begin{equation}
    \sum_{s=-\infty}^{\infty} i s A_s(\bar{d}) \, e^{is \bar{x}}  =  0,
    \label{eq:s_sA}
\end{equation}
where we have defined the functions $A_s(\bar{d})$ as
\begin{align}
    A_s(\bar{d})  &=  \sum_{\substack{r={-\infty} \\ r\equiv s (\mathrm{mod}\,2)}}^{\infty} a_{(r-s)/2}^* \, a_{(r+s)/2} 
    \\ &\times 
    \left[e^{-i (r-s)\bar{d}/2}-1\right]\left[e^{i (r+s)\bar{d}/2}-1\right]  .
\end{align}
The functions $\exp(is\bar{x})$ are linearly independent on $[0,2\pi]$.
Therefore, Eq.~\eqref{eq:s_sA} implies
\begin{equation}
    A_s(\bar{d}) =  0 \; \; \; \forall s\neq 0  ,
\end{equation}
while, because of the prefactor $s$ on the right-hand side of Eq.~\eqref{eq:s_sr}, no condition is imposed on $A_0(\bar{d})$. 
Let us, therefore, first consider the case $s \neq 0$. We have
\begin{align}
\nonumber
    & A_s(\bar{d}) = 
    \sum_{\substack{r={-\infty} \\ r\equiv s (\mathrm{mod}\,2)}}^{\infty}
    a_{(r-s)/2}^* a_{(r+s)/2} 
    \\ & \times \left[1+e^{i s\bar{d}} - e^{-\frac{i(r-s)\bar{d}}{2}} - e^{-\frac{i(r+s)\bar{d}}{2}}\right] = 0  \; \; \; \forall s\neq 0  .
    \label{eq:Asd1}
\end{align}
The coefficients $A_s(\bar{d})$ are identically zero if either $r-s = 0$ or $r+s = 0$.
This implies that the coefficient $a_0$ is arbitrary, regardless of the values of all the other coefficients in the sum.
This point will be discussed in more detail later.

The four terms within square brackets in Eq.~\eqref{eq:Asd1} are of the form $e^{ijd}$, where $j$ 
is an integer given by $0$, $s$, $(r-s)/2$, and $(r+s)/2$, respectively.
By collecting the terms corresponding to a given value of $j$, Eq.~\eqref{eq:Asd1} can be recast as
\begin{equation}
    A_s(\bar{d}) = \sum_{j=-\infty}^\infty B_{s,j} e^{ij\bar{d}}  = 0 \; \; \; \forall s\neq 0  ,
    \label{eq:Asd3}
\end{equation}
where we have defined
\begin{align}
\nonumber
    B_{s,0} &= B_{s,s} = \sum_{\substack{r={-\infty} \\ r\equiv s (\mathrm{mod}\,2)}}^{\infty}
    a_{(r-s)/2}^* a_{(r+s)/2} - a^*_0 a_s  - a^*_{-s} a_0
    \\ & = \sum_{\substack{r=-\infty \\ r \neq s,-s \\ r\equiv s (\mathrm{mod}\,2)}}^\infty a^*_{(r-s)/2} a_{(r+s)/2} 
    \label{eq:Bss}
\end{align}
and
\begin{equation}
    B_{s,j}  =  -a_{-j}^* \, a_{-j+s} \, - \, a_{j-s}^* \, a_{j} \; \; \; {\rm for} \; j\neq 0, s .
    \label{Eqn:bsj}
\end{equation}
Since the functions $e^{ij\bar{d}}$ are linearly independent on $[0,2\pi]$
, the only way to satisfy Eq.~\eqref{eq:Asd3} is to have 
\begin{equation}
    B_{s,j} = 0 \; \; \;  \forall s\neq0 \; \; \; {\rm and} \; \; \; \forall j  .
    \label{Eqn:bsj0}
\end{equation}

We will now prove that, besides $a_0$, the Fourier series of $q_L(x)$ contains a single non-zero coefficient $a_j$.
This is the main result of the present work, that has deep implications on the possible form of the periodic position $q_L(x)$.
In order to prove this fact we consider the case $j\neq 0$ and $j\neq s$ and we assume that 
$a_{\bar k} \neq 0$ for some specific $\bar k \neq 0$. We will then proceed in two steps:
\begin{enumerate}
    \item
    We first consider the case $s=2\bar k$ $(\bar{k} \neq 0)$. 
    From Eqs.~\eqref{Eqn:bsj} and \eqref{Eqn:bsj0} we then obtain 
    \begin{equation}
        B_{2\bar k,\bar k}  =  
        -2 a^*_{-\bar k} a_{\bar k} = 0
    \end{equation}
    %
    Since we assumed that $a_{\bar k} \neq 0$, we must have that $a_{-\bar k} = 0$.
    \item
    We then consider the case $s=\bar k - \bar l$ $(\bar{k} \neq \bar{l})$.
    From Eqs.~\eqref{Eqn:bsj} and \eqref{Eqn:bsj0} we then obtain 
    \begin{equation}
        B_{\bar k - \bar l,\bar k }  =  -a^*_{-\bar k} a_{-\bar l} - a^*_{\bar l} a_{\bar k}  =  0.
        \label{eq:Bsk}
    \end{equation}
    The first term on the right-hand side of Eq.~\eqref{eq:Bsk} is zero, since $a_{-\bar k}$ is zero because of point 1.
    Then, since we assumed $a_{\bar k} \neq 0$, we must have that $a_{\bar l} = 0$.
\end{enumerate}
This completes the proof.
Only a single exponential function $e^{ik\bar{x}}$ can appear in Fourier series of $q_L(x)$
given in Eq.~\eqref{Eqn:Fourier}.
Moreover, Condition 2 (simple periodicity) implies that this unique non-zero term will be associated either to $k=1$ or to $k=-1$.
Without loss of generality, we will use $k=1$ in the following.
In summary, we have shown that the only functions $q_L(x)$ that satisfy Conditions 1, 2 and 3 are
\begin{equation}
    q_L(x)  =  a_1 \exp\left(\frac{2\pi i}{L} x\right) + a_0
    \label{eq:qpm}
\end{equation}
where $a_0$ and $a_1$ are complex numbers.
We note that this implies that $q_L(x)$ is ${\cal C}^\infty$.

We will now take into account Condition 4 to determine the coefficient $a_1$.
The square distance corresponding to the position in Eq.~\eqref{eq:qpm} is given by
\begin{align}
    d^2_{12}  &=  |q_L(x_1)-q_L(x_2)|^2  
    \\ &=  
    4|a_1|^2 \sin^2 \left(\frac{\pi (x_1-x_2)}{L}\right)  .
\end{align}
A Taylor expansion up to first order of the above expression gives, for $L\gg |x_1-x_2|$,
\begin{equation}
    d^2_{12}  =  4|a_1|^2 \left(\frac{\pi^2 (x_1-x_2)^2}{L^2}\right) .
    \label{eq:dd12_prime}
\end{equation}
Condition 4 imposes that, in the limit $|x_1-x_2|/L \rightarrow 0$, 
we must have $d^2_{12}  =  (x_1-x_2)^2$.
We conclude that $a_1  =  \frac{L}{2\pi}e^{i\phi}$ with $\phi \in \cal{R}$.
Therefore we have
\begin{equation}
    q_L(x)  =  e^{i\phi}\frac{L}{2\pi}\exp\left(\frac{2\pi i}{L} x\right) + a_0.
    \label{Eqn:qlxa0}
\end{equation}
We thus obtain
%
%
\begin{equation}
    d^2_{12}  = \frac{L^2}{\pi^2} \sin^2 \left(\frac{\pi (x_1-x_2)}{L}\right)
    \label{eq:dd12}
\end{equation}
for the square distance which is purely real and independent of $a_0$ and $\phi$.

%
%
No physical constraint can be used to fix the value of $a_0$, 
which is a gauge parameter related to the choice for the origin of the coordinate system in the periodic system.
%
Therefore, for the sake of simplicity both $C$ and $\phi$ could be set to zero which would yield
\begin{equation}
    q_L(x)  =  \frac{L}{2\pi}\exp\left(\frac{2\pi i}{L} x\right).
    \label{Eqn:def1}
\end{equation}
An advantage of this expression is that it is characterized by a constant modulus, \emph{i.e.}, $|q_L(x)|  =  \frac{L}{2\pi}$.
Therefore, all points are equivalent.
It was used, for example, in Ref.~[\onlinecite{Angeli_2021}].
Alternatively, we can impose a stronger condition than that given in Eq.~\eqref{Eqn:cond4}.
Instead of imposing a constraint on the distance we can impose a similar constraint on the position according to
%
%
\begin{equation}
    \lim_{L \rightarrow \infty} q_L(x) =  x.
    \label{Eqn:cond5}
\end{equation}
In other words this constraint guarantees that (for a fixed value of $x$) when $L\rightarrow\infty$ the periodic position tends to the non-periodic position. 
A Taylor expansion around $x/L = 0$ up to first order on the right-hand side of Eq.~\eqref{Eqn:qlxa0} yields
\begin{equation}
    q_L(x)  =  a_0 + e^{i\phi}\left[ \frac{L}{2\pi} + ix \right] + O\left(\frac{x}{L}\right)^2
\end{equation}
Applying the constraint in Eq.~\eqref{Eqn:cond5} then yields $a_0 = -e^{i\phi}\frac{L}{2\pi}$ and $e^{i\phi} = -i$.
With this additional constraint the periodic position is thus given by
\begin{equation}
    q_L(x)  =  \frac{L}{2\pi i}\left[\exp\left(\frac{2\pi i}{L} x\right) - 1\right].
    \label{Eqn:def2}
\end{equation}
%



The two expressions given in Eqs.~\eqref{Eqn:def1} and \eqref{Eqn:def2}, as well as the
general expression in Eq.~\eqref{Eqn:qlxa0} all have in common the complex dependence on $x$.
Therefore, a periodic position that fulfills conditions 1-4 is intrinsically complex.
Although this could seem surprising, one can argue that 
what one can really measure are {\em distances} and not absolute positions, and as shown in Eq.~\eqref{eq:dd12}
the distances corresponding to the complex position operator are purely real.

Finally, we note that the function $q_L(x)$ should induce a metric on a space subject to PBC.
In other words, we require that, if we define the distance $d_{12}$ between $x_1$ and $x_2$ as $d_{12} \, \equiv \, |q_L(x_1)-q_L(x_2)|$, then $d_{12}$ should satisfy the three fundamental axioms of the distance in a metric space.
Since the restriction of a metric space to an arbitrary subset is still a metric space, the function $q_L(x)$ indeed induces a proper distance.
More specifically, we can see that the expression of the square distance between two points in  Eq.~\eqref{eq:dd12} is nothing but the Euclidean square distance between the points $P_1$ and $P_2$~, whose coordinates are given by $P_i  =  \Bigl(\frac{L}{2\pi} \cos(2\pi x_j/L) \, , \, \frac{L}{2\pi} \sin(2\pi x_j/L) \Bigr)$, with $j=1,2$.
Indeed, it is easy to show that $|P_1-P_2|=d_{12}$ .
Therefore $d_{12}$ is the restriction of the usual Euclidean distance on $\mathbb{R}^2$ to a circle of radius $R=L/(2\pi)$, 
and whose points $P_1$ and $P_2$ are identified by the arc lengths $x_1$ and $x_2$~.

The above proof in one dimension can be easily generalized to two and three dimensions, by applying the four conditions we imposed on the one-dimensional periodic position to each Cartesian component.
For instance, the periodic position in three dimensions is given by
\begin{equation}
    q_{L_x,L_y,L_z}(\vec{r}) = 	
    \frac{1}{2\pi i}
    \begin{pmatrix}
    	\vspace{0.1cm}
		L_x [\exp\left(\frac{2\pi i}{L_x} x\right) - 1]	\\
		\vspace{0.1cm}
		L_y [\exp\left(\frac{2\pi i}{L_y} y\right) - 1]	\\
		L_z [\exp\left(\frac{2\pi i}{L_z} z\right) - 1].
	\end{pmatrix}
\end{equation}
%
In the supplemental material we prove that the position defined in the above equation induces a metric, \emph{i.e.,} an acceptable distance between two points.

In conclusion, we have proven that, except for a phase factor and a constant, 
there exists a unique position and corresponding one-body position operator that is compatible with periodic boundary conditions.
This position operator, is inherently complex but the corresponding distance is purely real, as it should be.
The proof relies on the periodicity of the position and on the application of several physical constraints on the distance.
Finally, we showed that with an additional constraint on the position itself we can also fix the phase factor and the constant.
As mentioned before, our position operator has already been shown to be very useful to calculate various properties of periodic systems, such as Madelung constants, localization tensors, polarizabilities, etc. 
Any observable of a periodic system that can be related to the position operator could be calculated with our approach.

F.~A.~S.~ would like to thank  L’Oreal UNESCO “For Women in Science 2018” for partially supporting  this work.
We also thank Prof. Richard E. Schwartz for very helpful discussions and suggestions. 
J.~A.~B.~ thanks the French Agence Nationale de la Recherche (ANR) for financial support (Grant Agreement ANR-19-CE30-0011).
This work was partially supported by the “Programme Investissements d’Avenir” under the program ANR-11-IDEX- 0002-02, reference ANR-10-LABX-0037-NEXT.





%

\end{document}



\title{Supplemental material for "A unique one-body position operator for periodic systems"}

\author{Stefano Evangelisti}
\affiliation{\lcpq}
\author{Faten Abu-Shoga}
\affiliation{\uniga}
\author{Celestino Angeli}
\affiliation{\unife}
\author{Gian Luigi Bendazzoli}
\affiliation{\unibo}
\author{J.~Arjan Berger}
\affiliation{\lcpq}
\affiliation{\etsf}

\maketitle

\begin{thm}
 Suppose that $\left(X, d_X\right)$ and $\left(Y, d_Y\right)$ are metric spaces. Define the function $d_{XY}:\left(X \times Y\right) \times\left(X \times Y\right) \rightarrow \mathbb{R}$ by
$d_{XY}\left(\left\langle x_{1}, y_{1}\right\rangle,\left\langle x_{2}, y_{2}\right\rangle\right)=\sqrt{d_X\left(x_{1}, x_{2}\right)^{2}+d_Y\left(y_{1}, y_{2}\right)^{2}}$. Then $d_{XY}$ is a metric on $X \times Y$
\end{thm}

\begin{proof}
Firstly, $d_{XY}\left(\left\langle x_{1}, y_{1}\right\rangle,\left\langle x_{2}, y_{2}\right\rangle\right) \geq 0$ for all $\left\langle x_{1}, y_{1}\right\rangle$ and $\left\langle x_{2}, y_{2}\right\rangle$ because a square root is always
non-negative.
 We now show that $d_{XY}$ satisfies the three conditions of a metric space.
 
\textbf{Condition  1} Let $$d_{XY}\left(\left\langle x_{1}, y_{1}\right\rangle,\left\langle x_{2}, y_{2}\right\rangle\right)=\sqrt{d_X\left(x_{1}, x_{2}\right)^{2}+d_Y\left(y_{1}, y_{2}\right)^{2}}=0$$ 
this is true if and only if $d_X\left(x_{1}, x_{2}\right)=0$ and $d_Y\left(y_{1}, y_{2}\right)=0$, which is in turn true if and only if $x_{1}=x_{2}$ and $y_{1}=y_{2}$ because $d_X$ and $d_Y$ are metrics. More over , $x_{1}=x_{2}$ and $y_{1}=y_{2}$ if and only if $\left\langle x_{1}, y_{1}\right\rangle=\left\langle x_{2}, y_{2}\right\rangle$ . 

\textbf{Condition  2}, By using the symmetry of $(X,d_X)$ and $(Y,d_Y)$ we have 
$$d_{XY}\left(\left\langle x_{1}, y_{1}\right\rangle,\left\langle x_{2}, y_{2}\right\rangle\right)=\sqrt{d_X\left(x_{1}, x_{2}\right)^{2}+d_Y\left(y_{1}, y_{2}\right)^{2}}=\sqrt{d_X\left(x_{2}, x_{1}\right)^{2}+d_Y\left(y_{2}, y_{1}\right)^{2}}$$
$$=d_{XY}\left(\left\langle x_{2}, y_{2}\right\rangle,\left\langle x_{1}, y_{1}\right\rangle\right)$$

\textbf{Condition  3}, let $\left\langle x_{1}, y_{1}\right\rangle,\left\langle x_{2}, y_{2}\right\rangle$ and $\left\langle x_{3}, y_{3}\right\rangle$ be elements of $X \times Y$. From the triangle inequality for $d_X$, we know that $d_X\left(x_{1}, x_{3}\right) \leq d_X\left(x_{1}, x_{2}\right)+d_X\left(x_{2}, x_{3}\right) .$ Squaring both sides, we get
$$
\begin{aligned}
d_X\left(x_{1}, x_{3}\right)^{2} & \leq\left(d_X\left(x_{1}, x_{2}\right)+d_X\left(x_{2}, x_{3}\right)\right)^{2} \\
&=d_X\left(x_{1}, x_{2}\right)^{2}+2 d_X\left(x_{1}, x_{2}\right) d_X\left(x_{2}, x_{3}\right)+d_X\left(x_{2}, x_{3}\right)^{2} \\
& \leq d_X\left(x_{1}, x_{2}\right)^{2}+d_X\left(x_{2}, x_{3}\right)^{2}
\end{aligned}
$$

From the triangle inequality for $d_Y$, we know that $d_Y\left(y_{1}, y_{3}\right) \leq d_Y\left(y_{1}, y_{2}\right)+d_Y\left(y_{2}, y_{3}\right) .$ Squaring both sides, we get
$$
\begin{aligned}
d_Y\left(y_{1}, y_{3}\right)^{2} & \leq\left(d_Y\left(y_{1}, y_{2}\right)+d_Y\left(y_{2}, y_{3}\right)\right)^{2} \\
&=d_Y\left(y_{1}, y_{2}\right)^{2}+2 d_Y\left(y_{1}, y_{2}\right) d_Y\left(y_{2}, y_{3}\right)+d_Y\left(y_{2}, y_{3}\right)^{2} \\
& \leq d_Y\left(y_{1}, y_{2}\right)^{2}+d_Y\left(y_{2}, y_{3}\right)^{2}
\end{aligned}
$$
Adding these two inequalities  gives
$$
\begin{aligned}
&d_{XY}\left(\left\langle x_{1}, y_{1}\right\rangle,\left\langle x_{3}, y_{3}\right\rangle\right)^2 \\
&\quad=d_X\left(x_{1}, x_{3}\right)^{2}+d_Y\left(y_{1}, y_{3}\right)^{2} \\
&\quad \leq \left(d_X\left(x_{1}, x_{2}\right)^{2}+d_X\left(x_{2}, x_{3}\right)^{2}\right)+\left(\left(d_Y\left(y_{1}, y_{2}\right)^2+d_Y\left(y_{2}, y_{3}\right)\right)^{2}\right) \\
&\quad=\left(d_X\left(x_{1}, x_{2}\right)^{2}+d_Y\left(y_{1}, y_{2}\right)^{2}\right)+\left(d_X\left(x_{2}, x_{3}\right)^{2}+d_Y\left(y_{2}, y_{3}\right)^2 \right) \\
&\quad \leq d_X\left(x_{1}, x_{2}\right)^{2}+d_Y\left(y_{1}, y_{2}\right)^{2}+d_X\left(x_{2}, x_{3}\right)^{2}+d_Y\left(y_{2}, y_{3}\right)^2  \\
&\quad=d_{XY}\left(\left\langle x_{1}, y_{1}\right\rangle,\left\langle x_{2}, y_{2}\right\rangle\right)^2+d_{XY}\left(\left\langle x_{2}, y_{2}\right\rangle,\left\langle x_{3}, y_{3}\right\rangle\right)^2
\end{aligned}
$$
by  taking the square root and applying the fact that $$ \forall a,b \geq 0 ,~~~~\sqrt{a^2+b^2} \leq a+b $$  we have 

$$
\begin{aligned}
d_{XY}\left(\left\langle x_{1}, y_{1}\right\rangle,\left\langle x_{3}, y_{3}\right\rangle\right)&\leq  \sqrt{ d_{XY}\left(\left\langle x_{1}, y_{1}\right\rangle,\left\langle x_{2}, y_{2}\right\rangle\right)^2+d_{XY}\left(\left\langle x_{2}, y_{2}\right\rangle,\left\langle x_{3}, y_{3}\right\rangle\right)^2}\\
&\leq  d_{XY}\left(\left\langle x_{1}, y_{1}\right\rangle,\left\langle x_{2}, y_{2}\right\rangle\right)+d_{XY}\left(\left\langle x_{2}, y_{2}\right\rangle,\left\langle x_{3}, y_{3}\right\rangle\right)
\end{aligned} $$
\end{proof}